\documentclass[conference]{IEEEtran}
\IEEEoverridecommandlockouts
\usepackage{cite}
\usepackage{amsmath,amssymb,amsfonts}
\usepackage{algorithmic}
\usepackage{graphicx}
\usepackage{textcomp}
\usepackage{xcolor}
\usepackage[explicit]{titlesec}
\usepackage{ulem}
\usepackage{lipsum}
\usepackage{multirow}
\usepackage{hyperref}
\usepackage{caption}
\usepackage{float}
\def\BibTeX{{\rm B\kern-.05em{\sc i\kern-.025em b}\kern-.08em
    T\kern-.1667em\lower.7ex\hbox{E}\kern-.125emX}}
    
\begin{document}

\title{MVD: A Novel Methodology and Dataset for Acoustic Vehicle Type Classification\\
}

\makeatletter
\newcommand{\linebreakand}{%
  \end{@IEEEauthorhalign}
  \hfill\mbox{}\par
  \mbox{}\hfill\begin{@IEEEauthorhalign}
}
\makeatother
\author{\IEEEauthorblockN{Mohd Ashhad}
\IEEEauthorblockA{\textit{Dept. of Computer Science} \\
\textit{Jamia Hamdard}\\
Delhi, India \\
ashhad.jamiahamdard@gmail.com}
\and
\IEEEauthorblockN{Omar Ahmed}
\IEEEauthorblockA{\textit{Dept. of Computer science} \\
\textit{Jamia Hamdard}\\
Delhi, India \\
ahmed.oas.omar@gmail.com}
\and
\IEEEauthorblockN{Sooraj K. Ambat}
\IEEEauthorblockA{\textit{Naval, Physical \& Oceanographic Laboratory} \\
\textit{Defence R\&D Organization}\\
Kochi, India \\
sooraj.npol@gov.in}
\linebreakand
\IEEEauthorblockN{Zeeshan Ali Haq}
\IEEEauthorblockA{\textit{Dept. of Computer Science} \\
\textit{Jamia Hamdard}\\
Delhi, India \\
aliz.inam@gmail.com}
\and
\IEEEauthorblockN{Mansaf Alam}
\IEEEauthorblockA{\textit{Dept. of Computer Science} \\
\textit{Jamia Millia Islamia}\\
Delhi, India \\
malam2@jmi.ac.in}
}

\maketitle

\begin{abstract}
Rising urban populations have led to a surge in vehicle use and made traffic monitoring and management indispensable. Acoustic traffic monitoring (ATM) offers a cost-effective and efficient alternative to more computationally expensive methods of monitoring traffic such as those involving computer vision technologies. In this paper, we present MVD and MVDA: two open datasets for the development of acoustic traffic monitoring and vehicle-type classification algorithms, which contain audio recordings of moving vehicles. The dataset contain four classes— Trucks, Cars, Motorbikes, and a No-vehicle class. Additionally, we propose a novel and efficient way to accurately classify these acoustic signals using cepstrum and spectrum based local and global audio features, and a multi-input neural network. Experimental results show that our methodology improves upon the established baselines of previous works and achieves an accuracy of 91.98\% and 96.66\% on MVD and MVDA Datasets, respectively. Finally, the proposed model was deployed through an Android application to make it accessible for testing and demonstrate its efficacy.

\end{abstract}

\section{Introduction}
Increasing urban city populations have made automated traffic management vital to avoid traffic jams, and road accidents and promote road safety\cite{Masum2018}\cite{Rafiq2021}. With an increasing need for smart cities, traffic monitoring has received a lot of attention from researchers in the past decade\cite{Pellicer2013} Traffic monitoring techniques can be used for various applications such as vehicle detection, sub-type classification, speed estimation, traffic density estimation, prediction of road accidents,  traffic signal optimization, etc. Acoustic Traffic Monitoring (ATM) algorithms have shown to be better alternatives to vision-based approaches as they are more computationally efficient\cite{Lefebvre2017}, not sensitive to lighting conditions\cite{Tyagi2012}, easily deployable, and have the added benefit of preserving the privacy of drivers. 

ATM algorithms come with their own set of challenges. Vehicle detection and type classification using just the sound signature is a challenging task as the sound is a mixture of engine sounds, sounds of tires (which may vary due to air pressure), road conditions, and environmental sounds such as wind, rain, horns, pedestrians, animals, etc. Moreover, the lack of large, high-quality, balanced datasets has made the development of ATM algorithms a challenging task which often compels researchers to record their own samples and makes the validation and comparison of algorithms harder.

To address the challenges associated with the lack of open datasets and to improve the performance of acoustic vehicle type classification, we introduce two datasets, namely MVD and MVDA, which comprise audio recordings of four classes of vehicles and a wide range of ambient noises. We've devised a robust approach for identifying vehicle sounds by leveraging both global and local audio features combined with a multi-input neural network. This approach outperforms existing models, all while employing significantly fewer trainable parameters. Additionally, we've developed an Android application to assess our trained model's performance seamlessly.
The main contributions of our paper are as follows:

\begin{itemize}

    \item We present the MVD (\textbf{M}oving \textbf{V}ehicle \textbf{D}etection) and MVDA (\textbf{M}oving \textbf{V}ehicle \textbf{D}etection \textbf{A}ugmented) Datasets. The former contains 4229 recordings of trucks, cars, motorbikes, and a no-vehicle class, which includes recordings of diverse background noises. The latter is a larger augmented version containing 16,916 recordings for better development and evaluation of ATM algorithms.

    \item We propose a robust methodology for acoustic vehicle type classification using global and local audio features and a multi-input neural network. Our methodology achieves an accuracy of $91.98\%$ and $96.66\%$ on the MVD and MVDA datasets respectively and improves upon the baseline performance on the IDMT-Traffic\cite{Abesser2021} and IDMT-Traffic$^\dagger$\cite{Ashhad2023} datasets while using $97.87\%$ and $95.65\%$ fewer trainable parameters respectively. 
    
    \item Finally, we deploy our trained model through an Android application to assess the performance of the model and investigate its dependence on the quality of the microphone. The app is freely available for testing.

\end{itemize}

 The structure of the paper is as follows: Section II of the paper provides a brief overview of the recent ATM research. Section III describes the MVD, MVDA, and other datasets utilized for the study. The proposed methodology for the work is discussed in Section IV. The results of the various experiments are presented in Section V. Finally, Section VI concludes the paper.
 
\section{Previous work}
Traffic monitoring has a wide range of applications. However, in this paper, we will limit our discussion to vehicle type classification. Over the past decade, researchers have used many methods for this task. Broadly, these methods can be classified into three groups: computer vision-based approaches, seismic monitoring, which uses sensors to monitor the ground vibrations; and finally, acoustic methods that utilize the sound of passing vehicles.

Sarikan et al.\cite{Sarikan2017}  achieved almost perfect classification on a dataset of 1200 samples of cars and motorbikes using a decision tree and k-nearest neighbor. They captured information from a light curtain and used circularity, heat map, and skeleton as features for the classifier. Chen et al.\cite{Chen2011} proposed a feature vector of thirteen measurements, which were subjected to Random Forest and Support Vector Machine classifiers. The authors utilized 2055 CCTV images and reported an accuracy of 96.26\%. Bhujbal et al.\cite{Bhujbal2020} used convolutional neural network layers as feature extractors to generate labels on 2659 images obtained from cityscapes. Using the YOLO technique, authors achieved 87\% accuracy. Sang et al.\cite{Sang2018} achieved a mean Average Precision (mAP) of 94.78\% on the BIT-Vehicle dataset. A revamped version of YOLOv2 called YOLOv2\_Vehicle and k-means++ clustering algorithm was engaged to obtain better anchor boxes, feature extraction, and weather adaptability.
   
Jin et al.\cite{Jin2018} extracted log-scaled frequency cepstral coefficients (LFCC) from seismic recordings of DARPA’s SITEX02 Dataset. They achieved an accuracy of 91.93\%. Kalra et al.\cite{Kalra2020}, recorded seismic signals of vehicles as well as noise using geophones and used Empirical Wavelet Transform (EWT) based time-frequency coefficients to extract statistical features such as mean, kurtosis, Renyi entropy, etc., and used an SVM based classifier to get a F-Score of  78\%, 67\%, and 86\% for bus, tractor, and noise, respectively. Bin et al.\cite{Bin2021} used Compressed Sensing (CS) theory to compress seismic signals and used a CNN to achieve on-site efficient classification of military vehicles. They managed to achieve an accuracy of 92.05\% on the SITEX02 dataset. Later, Bin et al.\cite{Bin2021a} used Fractal Theory to extract nonlinear seismic features and used SVM as a classifier to achieve an F1 score of 0.901 on the same dataset.
 
Liu et al.\cite{Liu2022} proposed a distributed multi-class Gaussian process (MCGP) algorithm for vehicle type classification. Their experimentation on the acoustic-seismic classification identification data set (ACIDS) revealed a test error between 0.3 and 0.06. Mohine et al.\cite{Mohine2022} proposed a hybrid convolutional neural network-bidirectional long short-term memory model (CNN-BiLSTM) that extracts features from acoustic signals and classifies the acoustic signal into five classes (two-wheeler, low, medium and heavy weight vehicles and noise). The authors achieved 92\% accuracy on their custom dataset and 96\% on the SITEX02 acoustic dataset. Ashhad et al.\cite{Ashhad2023} attained 98.95\% accuracy on a carefully pre-processed version of the IDMT Traffic\cite{Abesser2021} dataset by deploying an efficient CNN with MFCC features. Chen et al.\cite{Chen2021} proposed a hybrid neural network classifier with long short-term memory (LSTM) units fused into CNN layers. The authors extracted MFCC, pitch class profile (PCP) and short-term energy (STE) features and reported an accuracy of 97.65\% on the SITEX02 acoustic dataset.

\section{Datasets}

\subsection{MVD Dataset}
In this paper, we present two novel datasets. The MVD dataset consists of 4229 audio samples of passing vehicles moving at variable speeds, recorded with four separate high-quality MEMS mono channel microphones at four different urban locations in the Delhi-NCR region of India. The recording scenarios consist of wet and dry road conditions as well as a plethora of environmental noise such as wind, rain, animals, pedestrians and vehicles playing loud music. The samples were captured at a sampling rate of 22,050 kHz. Each audio is padded to be the same length of 3 seconds. The dataset includes four classes: cars (1005 samples), trucks (1077 samples), motorcycles (1122 samples), and a no-vehicle class (1025 samples).

\subsection{MVDA Dataset}
Apart from the MVD we also present the MVDA (Moving Vehicle Detection Augmented) dataset which is an augmented version of the MVD dataset. The motivation behind building this variation of the dataset lies in the fact that exposing a machine learning model to a gamut of sound variations, and audio augmentations improves the model's ability to generalize. This means that the model is better able to recognize similar sounds, even if they are not exact replicas of the sounds in the training dataset. The MVDA dataset contains all the recordings of the MVD dataset as well as augmented variants of each of these recordings with the following three augmentations:
\vspace{0.3cm}
\begin{itemize}

    \item \textbf{Random gain:} This technique involves applying a random gain (i.e., amplification or attenuation) to the audio signal. This is done by multiplying the audio signal by a random gain factor between a specified range as illustrated in the following equation: 
    \begin{equation}
   y(t)=x(t) \times g
   \end{equation}
   Here $y(t)$ is the output signal, $x(t)$ is the input signal, $t$ denotes time and $g$ is the random gain factor. We used minimum and maximum gain factors as 0.1 and 2 respectively. In practice, this augmentation has an effect similar to that of changing the distance of the vehicle from the microphone.
\vspace{0.3cm}
    \item \textbf{Noise injection:} In this technique, a random amount of white noise is added to the original audio signal to simulate various types of environmental or recording noise as illustrated in the following equation:
    \begin{equation}
   y(t)=x(t) + r\times n
   \end{equation}
    Here $y(t)$ is the output signal, $x(t)$ is the input signal, $n$ denotes Gaussian white noise and $r$ is the noise rate which is used to control the amount of noise added to the signal. For each sample, we randomly applied a noise rate between 0.001 and 0.003. This range was carefully fine-tuned so as to not allow the signal-to-noise ratio (SNR) to get too small. This augmentation was done to simulate the disturbance caused in recordings due to environmental noise.
\vspace{0.3cm}
    \item \textbf{Time stretching:} It is an audio augmentation technique that involves altering the speed or duration of an audio signal without changing its pitch as illustrated in the following equation.
    \begin{equation}
    y(t) = x(S \times t)
   \end{equation}
     Here $y(t)$ is the output signal, $x(t)$ is the input signal, $t$ denotes time and $S$ is the time scaling factor. For each sample in our dataset, we randomly applied a stretching factor between 0.8 and 1.5. In practice, this augmentation has an effect similar to that of changing the speed of the vehicle recorded.
   
\end{itemize}

\subsection{IDMT-Traffic Dataset}
In this work, we make use of the IDMT Traffic dataset\cite{Abesser2021}. Introduced in 2021, the IDMT Traffic dataset serves as an open benchmark dataset designed for acoustic traffic monitoring purposes. It encompasses stereo audio recordings that are time-synchronized, capturing the sounds of moving vehicles in four distinct recording locations. These locations include three urban traffic settings and one rural road location situated in and around Ilmenau, Germany. The recordings were obtained using two types of microphones: high-quality sE8 microphones and more affordable microelectromechanical systems (MEMS) microphones. The dataset covers various recording scenarios, encompassing different speed limits (30, 50, and 70 km/h) and varying road conditions, including both dry and wet surfaces. Specifically, the dataset comprises the following vehicle categories: Cars (3903 events), Trucks (511 events), Busses (53 events), and Motorcycles (251 events). Additionally, background recordings are available, representing situations when no vehicles are present.

\subsection{IDMT-Traffic$^{\dagger}$ Dataset}

Ashhad et al.\cite{Ashhad2023} suggested a possible problem with the truck samples of the IDMT-Traffic dataset leading to poor accuracy of the model in \cite{Abesser2021} on the truck class. They proposed a methodology for sample rejection and data augmentation for better training and evaluation of models trained on samples belonging to the IDMT-Traffic dataset. We refer to this version of the dataset as IDMT-Traffic$^\dagger$.

\section{Methodology}
In this section, we present the details of our method as illustrated in Fig. \ref{fig:METH}.

 \begin{figure}[hb]
\centerline{\includegraphics[width=9cm, height=11.5cm]{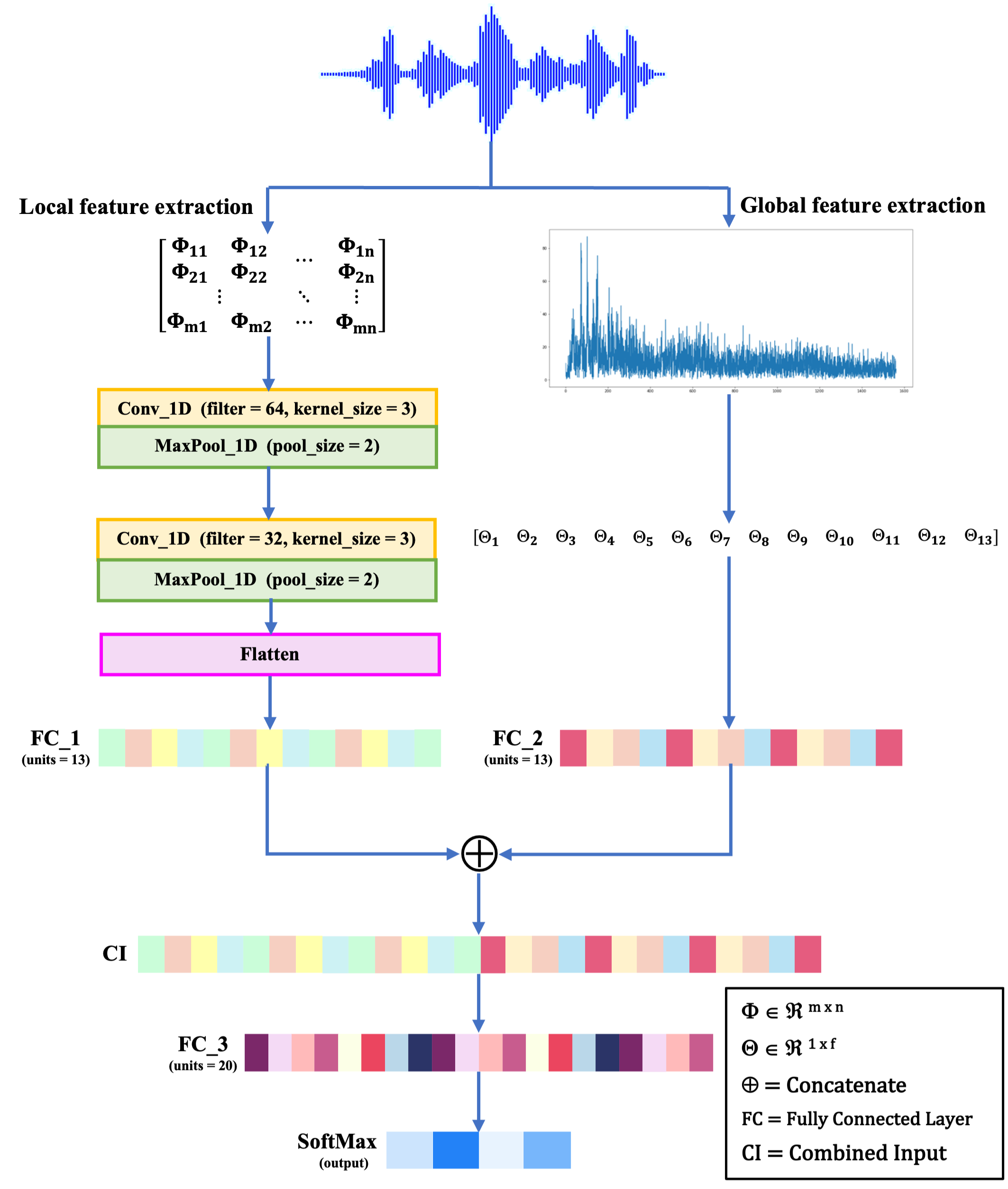}}
\centering
\captionsetup{justification=centering}
\caption{\small{Methodology of proposed model.}}
\label{fig:METH}
\end{figure}

\subsection{Feature Extraction and Audio Pre-processing}

 In our audio pre-processing pipeline, we implemented two crucial techniques: zero padding and pre-emphasis. These steps play a vital role in enhancing the quality and effectiveness of our audio data analysis. Our study involved the extraction of local and global audio features. We consider those features as local, which give a time localized or frame-wise estimation of the frequency, intensity, and energy contours of the audio samples, such as mel-spectrograms, MFCCs, GFCCs, etc., while global features convey information about the entire signal.

\vspace{0.3cm}
\subsubsection{Local audio features}
\vspace{0.3cm}
\begin{itemize}

    \item \textbf{Mel-spectrogram:} A mel-spectrogram is a mathematical representation of the spectrum of a sound signal, where the frequency axis is divided into mel frequency bands instead of linear frequency bands. To compute the mel-spectrogram we used a window size of 1024, a hop size of 512, and an FFT size of 2048. We used 128 mel-bands in the filter bank and applied log magnitude scaling.
    \vspace{0.3cm}
    \item \textbf{MFCC:} Mel Frequency Cepstrum Coefficients (MFCC) are a set of coefficients that represent the spectral envelope of the sound signal in the mel frequency domain. For our experiments, we utilized the first 40 coefficients while the rest of the parameters were the same as the previous section.
    \vspace{0.3cm}
   \item \textbf{GFCC:} After the success of the mel filterbank, Patterson and Smith\cite{Patterson1987} created a more comprehensive model that mimicked the features of the ear based on psychophysical research of the auditory periphery, called the Gammatone Filterbank. The human auditory systems are modelled by the Gammatone filter bank as a set of overlapping band-pass filters. The impulse response of each filter is given by the following equation:
   \begin{equation}
   g(t)=a{t}^{n-1}{e}^{-2\pi bt}cos(2 \pi f_ct+ \phi) 
   \end{equation}
    Here $t$ denotes time, $n$ is the order of the filter, $a$ is a constant (usually equal to 1), $b$ is the bandwidth parameter, $f_{c}$ is the filter center frequency and $\phi$ is the phase of the fine-structure of the impulse response.
    Based on the Gammatone filterbank, Xu et al.\cite{Xu2012} developed a feature extraction method called the Gammatone Frequency Cepstrum Coefficients (GFCC) which performed better than MFCC in certain tasks.
    
\end{itemize}
\vspace{0.3cm}
\subsubsection{Global audio features}
\vspace{0.3cm}
We extract statistical features from the spectrum of the audio signals as a second set of features since local features alone might be inadequate for robust acoustic vehicle classification. The motivation behind including these features in our model is that the pattern of these spectra can be used to infer specific differences in the frequency distribution and associated magnitudes among the sound signals of the various classes of vehicles. Additionally, the magnitude of the pure-tone frequencies in the spectrum may fluctuate as the distance between the car and the microphone varies, but their distribution (shape) remains similar, which can be used to increase the robustness of our model. Fig. \ref{Spec}(a),(b),(c) depicts the spectrums of motorcycles, cars, and trucks, respectively. It is clear from the figure that the shape of the spectrum of samples belonging to the same class is independent of the distance of the recording microphone from the vehicle. We calculate thirteen statistical features to create our input global feature vector. The features are as follows: kurtosis, skewness, standard deviation, variance, mode, iqr, mean, geometric mean, harmonic mean, median absolute deviation, variation, geometric standard deviation, and entropy.

\begin{figure}[htp]
\centerline{\includegraphics[width=9cm, height=6.5cm]{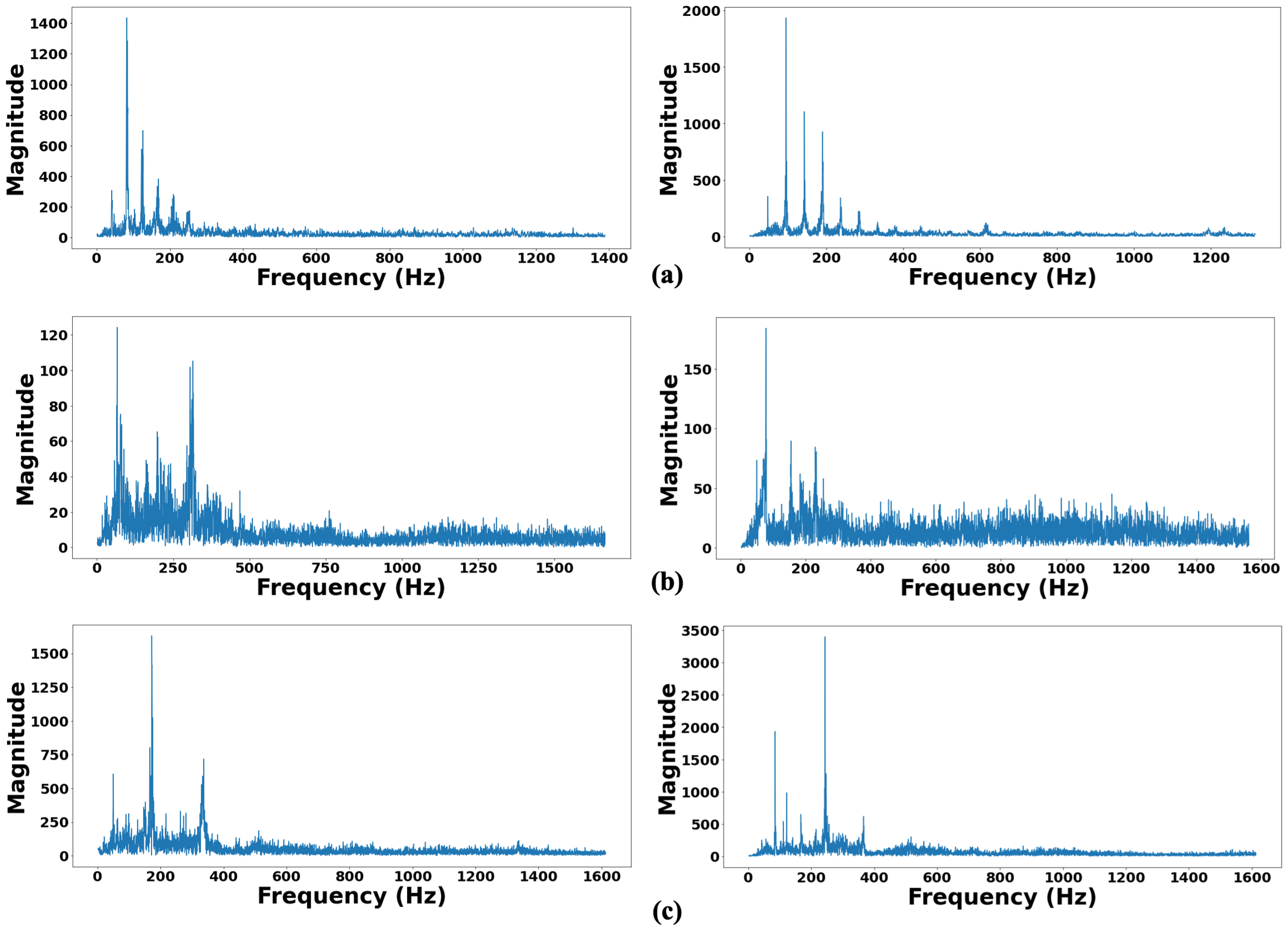}}
\centering
\captionsetup{justification=centering}
\caption{\small{Spectra of (a) bikes, (b) cars, and (c) trucks recorded at two-meter (left) and one-meter (right) illustrates similarities in signal envelopes within each class and the dissimilarities between the classes.}}
\label{Spec}
\end{figure}

\subsection{Neural Network Architecture}
We developed a multi-input neural network architecture that works well for the task after extensive testing and review.  Local feature matrix of size $m \times n$ (where $m$ corresponds to the frame index and $n$ to the index of cepstrum coefficient or frequency bin) is fed into 1D-CNN layers, while the global feature vector of size $1 \times f$ (where $f$ corresponds to the index of feature) is fed in parallel to a fully connected layer. The outputs of the parallel layers are then concatenated and passed to another fully connected layer. The final softmax layer gives the output label. Fig. \ref{fig:METH} illustrates our methodology and the architecture of the model. 

\begin{figure}[ht]
\centerline{\includegraphics[width=6cm, height=4cm]{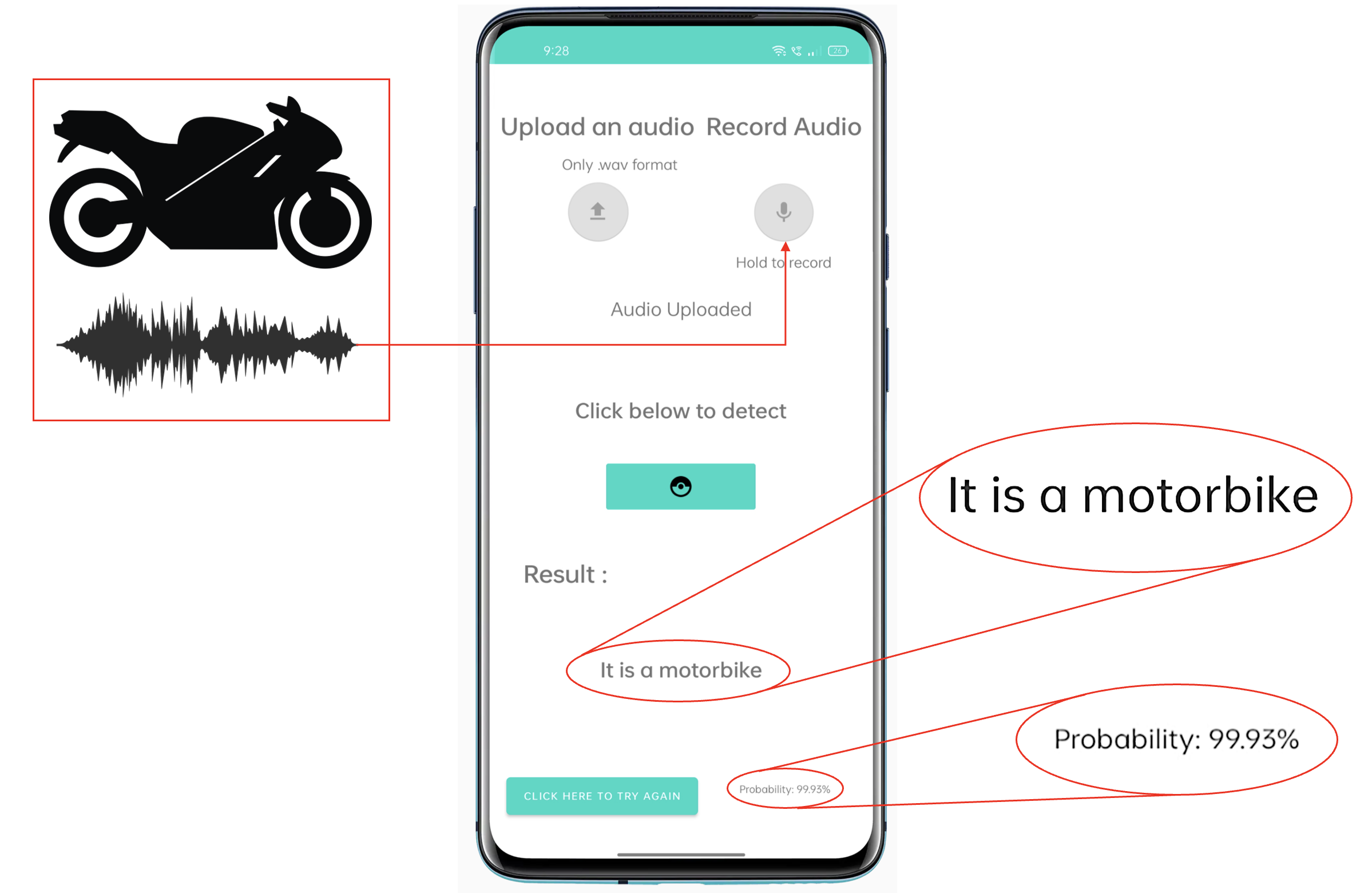}}
\centering
\captionsetup{justification=centering}
\caption{\small{Working of the MVD android application. The user can choose between recording a new sample or loading an existing one. The app then shows the prediction as well as the confidence score.}}
\label{fig:app}
\end{figure}

\subsection{Deployment through Android application}
We deployed our model through an Android application for mobile devices. The application is capable of capturing audio samples in real-time as well as loading existing audio files, extracting the proposed features from the audio, feeding the features into the trained model, and finally predicting the vehicle type (refer Fig. \ref{fig:app}). The application makes use of Chaquopy to run Python scripts. For prompt processing, the model is hosted on a cloud platform, where it receives the input from the app, processes it, and sends the output back through API calls. The App is developed in Kotlin and uses the Flask API for the backend.

 \section{Experiments and Results}
All the experiments were performed on Google Colab Pro with the Tensorflow framework in Python. All the models were trained for 50 epochs with early stopping and learning rate reduction with patience of 4 (minimum learning rate = 1e-3). We use the categorical cross-entropy loss function and Adam optimizer.
% \vspace{-0.3cm}

\subsection{Study of local features}
We conducted multiple experiments aimed at determining the optimal combination of global and local features for vehicle type classification. Understanding the most effective feature set was crucial for improving the accuracy and efficiency of our model. We experimented with MFCC, GFCC, and mel-spectrograms (MS) on the MVD dataset with 5-fold validation. As for the classifier, we use our fine-tuned multi-input neural network (refer Fig. \ref{fig:METH}). The study's findings are summarised in Table \ref{tab:mvd}. It is clear that GFCC, combined with global features produces the most accurate predictions with an accuracy of $91.98\%.$
\begin{table}[h]
    \caption{Classification results on the MVD dataset with varied features}
    \begin{center}
    \begin{tabular}{|p{1.3cm}|p{1.9cm}|p{1cm}|p{0.7cm}|p{0.5cm}|p{1cm}|}
\hline
Class & Feature & Precision & Recall & F1-Score & Accuracy (5-fold)\\ \hline

Car & \multirow{4}{*}{Global + MS} & 0.81 & 0.76 & 0.79 & \multirow{4}{*}{86.80\%}  \\ \cline{1-1} \cline{3-5} 
Motorcycle &  & 0.80 & 0.82 & 0.81 &   \\ \cline{1-1} \cline{3-5}
Truck &  & 0.92 & 0.93 & 0.93 &   \\ \cline{1-1} \cline{3-5}
No Vehicle &  & 0.96 & 0.92 & 0.94 &   \\ \cline{1-1} \cline{3-5} \hline

Car & \multirow{4}{*}{Global + MFCC} & 0.83 & 0.83 & 0.83 & \multirow{4}{*}{89.83\%} \\ \cline{1-1} \cline{3-5} 
Motorcycle &  & 0.89 & 0.96 & 0.93 &   \\ \cline{1-1} \cline{3-5} 
Truck &  & 0.86 & 0.83 & 0.84 &   \\ \cline{1-1} \cline{3-5}
No Vehicle &  & 1.00 & 0.99 & 0.99 &   \\ \cline{1-1} \cline{3-5} \hline

Car & \multirow{4}{*}{Global + GFCC} & 0.88 & 0.85 & 0.86 & \multirow{4}{*}{\textbf{91.98\%}}  \\ \cline{1-1} \cline{3-5} 
Motorcycle &  & 0.86 & 0.88 & 0.87 &   \\ \cline{1-1} \cline{3-5} 
Truck &  & 0.96 & 0.95 & 0.95 &   \\ \cline{1-1} \cline{3-5}
No Vehicle &  & 0.99 & 0.99 & 0.99 &   \\ \cline{1-1} \cline{3-5} \hline
\end{tabular}
\label{tab:mvd}
    \end{center}
    \end{table}
% \vspace{-0.3cm}
    
\subsection{Performance of the proposed model}
We evaluated the performance of our proposed methodology on 4 datasets: MVD, MVDA, IDMT Traffic\cite{Abesser2021} and IDMT-Traffic$^\dagger$\cite{Ashhad2023}. We performed 5-fold validation on the MVD and MVDA datasets (refer Table \ref{results}). The comparative results of our proposed model on the IDMT-Traffic and IDMT-Traffic$^\dagger$ datasets are shown in Table \ref{tab:baselines}. For a fair comparison, we use the same train-test split as employed by the authors in the original papers. It is clear from Table \ref{tab:baselines} that our proposed methodology improves upon the established baselines while using $97.87\%$ and $95.65\%$ fewer trainable parameters respectively (refer Table \ref{tab:params}).

\begin{table}[h]
    \caption{Performance of the proposed model on MVD \& MVDA datasets }
    \begin{center}
    \begin{tabular}{|p{1.3cm}|p{1.6cm}|p{1cm}|p{0.7cm}|p{0.5cm}|p{1cm}|}
\hline
Class & Dataset & Precision & Recall & F1-Score & Accuracy (5-fold)\\ \hline
Car & \multirow{4}{*}{MVD} & 0.88 & 0.85 & 0.86 & \multirow{4}{*}{91.98\%}  \\ \cline{1-1} \cline{3-5} 
Motorcycle &  & 0.86 & 0.88 & 0.87 &   \\ \cline{1-1} \cline{3-5} 
Truck &  & 0.96 & 0.95 & 0.95 &   \\ \cline{1-1} \cline{3-5}
No Vehicle &  & 0.99 & 0.99 & 0.99 &   \\ \cline{1-1} \cline{3-5} \hline

Car & \multirow{4}{*}{MVDA} & 0.95 & 0.95 & 0.95 & \multirow{4}{*}{96.66\%}  \\ \cline{1-1} \cline{3-5} 
Motorcycle &  & 0.96 & 0.96 & 0.96 &   \\ \cline{1-1} \cline{3-5} 
Truck &  & 0.97 & 0.96 & 0.97 &   \\ \cline{1-1} \cline{3-5} 
No Vehicle &  & 0.99 & 0.99 & 0.99 &   \\ \cline{1-1} \cline{3-5} \hline

\end{tabular}
\label{results}
    \end{center}
    \end{table}

\begin{table}[h]
    \caption{Class-wise F1 score of our model against baselines}
    \begin{center}
    \begin{tabular}{|p{1.2cm}|p{1.2cm}|p{0.7cm}|p{0.7cm}|p{1.2cm}|p{1.3cm}| }
     \hline
    %  \multicolumn{2}{|c|}{Comparison} \\
     Model& Dataset& Car & Truck & Motorcycle & No Vehicle \\
     \hline
     Abeßer, J et al.\cite{Abesser2021}  & IDMT-Traffic & 0.94 & 0.5 & 0.96 & 1.00 \\
     \hline
    \textbf{Proposed Model} & \textbf{IDMT-Traffic }&  \textbf{0.96} & \textbf{0.57} & \textbf{0.98} & \textbf{1.00} \\
     \hline
     Ashhad et al.\cite{Ashhad2023}  & IDMT-Traffic$^\dagger$ &   0.99 & 0.92 & 0.99 & 1.00 \\
     \hline

    \textbf{Proposed Model} & \textbf{IDMT-Traffic$^\dagger$} &  \textbf{0.99} & \textbf{0.94} & \textbf{0.99} & \textbf{1.00} \\
     \hline
    \end{tabular}
    \label{tab:baselines}
    \end{center}
    \end{table}
%\vspace{-0.3cm}

\begin{table}[b]
    \caption{Comparison of the number of trainable parameters of the proposed model with baselines}
    \begin{center}
    \begin{tabular}{|p{2.5cm}|p{2.5cm}|p{2.5cm}|}
     \hline
    %  \multicolumn{2}{|c|}{Comparison} \\
     Model& Features & \#Parameters \\
     \hline
     Abeßer, J et al.\cite{Abesser2021}  & Mel-spectrogram & 1,400,000 \\
     \hline
     Ashhad et al.\cite{Ashhad2023}  & MFCC &  712,436 \\
     \hline
    \textbf{Proposed Model} & GFCC + Statisical & 30,471 \\
     \hline
    \end{tabular}
    \end{center}
    \label{tab:params}
    \end{table}
    
\begin{table}[b]
    \caption{Confusion Matrix of predictions with Android app}
    \begin{center}
    \begin{tabular}{|p{1.8cm}|p{1.2cm}|p{1.1cm}|p{1.3cm}|p{1.3cm}| }
     \hline
    %  \multicolumn{2}{|c|}{Comparison} \\
     Class& Car & Truck & Motorcycle & No Vehicle \\
     \hline
     Car & 20 & 2 & 3 & 0 \\
     \hline
     Truck  &   3 & 23 & 1 & 0 \\
     \hline
     Motorcycle & 1& 0 &21 &0 \\
     \hline
    No vehicle &1 & 0 & 0 &25 \\
     \hline
    \end{tabular}
    \label{tab:confusion}
    \end{center}
    \end{table}

\subsection{Testing of the proposed model through Android application}
The existing models for acoustic vehicle type classification can be quite large for efficient training and deployment \cite{Abesser2021}\cite{Ashhad2023} \cite{Mohine2022}. Through our proposed model, we managed to reduce the complexity without compromising the performance. To demonstrate the efficiency of our model and study the dependence of its performance on the microphone quality, we deployed it as an application for Android devices. We tested the application on five separate smartphones. Each smartphone was used to record 5 samples of each class (20 samples per phone) at various urban locations, and then the application was used to perform predictions using the trained model. The results have been presented as a confusion matrix (refer to Table \ref{tab:confusion}). With an accuracy of $90\%$ our model performs well at the task irrespective of the quality of the microphone utilized for recording, thus increasing the utility of our approach for real-world conditions.

\section{Conclusion}
In this paper, we present the MVD and MVDA datasets, two new open datasets for the development and testing of ATM algorithms. With our proposed multi-input neural network and carefully extracted audio features, we managed to outperform baseline models at the task of acoustic vehicle type classification while using $95.65\%$ to  $97.87\%$ fewer trainable parameters. Additionally, through our Android application, we showed that our methodology is independent of the quality of the microphone and can be deployed with ease. In further research, we want to investigate the deployability of ATM algorithms on edge devices with limited computing capacity.
\vspace{0.1cm}
\newline
Reproducible Results: In the spirit of reproducible research, we provide necessary codes, data and android application publicly\cite{Ash2023}. 

 \section*{Acknowledgements}
The authors would like to thank Ashad Naushad (Dept. of CSE, Jamia Hamdard) for his invaluable contribution to the app development process.

\bibliographystyle{IEEE}
\bibliography{MVD}

\vspace{12pt}

\end{document}